\documentclass[aps,prb,twocolumn,superscriptaddress,floatfix]{revtex4-2}
\usepackage[utf8]{inputenc}
\usepackage[english]{babel}
\usepackage{amsmath}
\usepackage{upgreek} 
\usepackage{feynmf}
\usepackage{xcolor}
\usepackage{soul}
\usepackage{graphicx}
\usepackage{amsfonts}
\usepackage{blindtext}
\usepackage{appendix}
\usepackage[colorlinks=true,unicode=true,allcolors=blue]{hyperref}
\usepackage{siunitx}
\usepackage[normalem]{ulem}
\addto\captionsenglish{}

\begin{document}

\preprint{APS/123-QED}
\title{Optoacoustic cooling of traveling hypersound waves}
\author{Laura Blázquez Martínez}
\altaffiliation{Authors contributed equally to this work.}
\author{Philipp Wiedemann}%
\altaffiliation{Authors contributed equally to this work.}
\author{Changlong Zhu}
\altaffiliation{Authors contributed equally to this work.}
\author{Andreas Geilen}
\author{Birgit Stiller}
 \email[Corresponding author ]{birgit.stiller@mpl.mpg.de}
\affiliation{%
 Max Planck Institute for the Science of Light, Staudtstr. 2, 91058, Erlangen, Germany}%
 \affiliation{Department of Physics, Friedrich-Alexander Universität Erlangen-Nürnberg, Staudtstr. 7, 91058 Erlangen, Germany}
\date{\today}

\begin{abstract}
We experimentally demonstrate optoacoustic cooling via stimulated Brillouin-Mandelstam scattering in a 50\,cm-long tapered photonic crystal fiber. For a 7.38\,GHz acoustic mode, a cooling rate of 219\,K from room temperature has been achieved. As anti-Stokes and Stokes Brillouin processes naturally break the symmetry of phonon cooling and heating, resolved sideband schemes are not necessary. The experiments pave the way to explore the classical to quantum transition for macroscopic objects and could enable new quantum technologies in terms of storage and repeater schemes.
\end{abstract}

\maketitle

\section{Introduction} 
Cooling mechanical vibrations to the quantum ground state has recently been achieved in diverse optomechanical cavity configurations, such as micromechanical bulk resonators \cite{oconnellQuantumGroundState2010}, drums embedded in superconducting microwave circuits \cite{teufelSidebandCoolingMicromechanical2011}, silicon optomechanical nanocavities \cite{chanLaserCoolingNanomechanical2011}, levitated nanoparticles \cite{delicCoolingLevitatedNanoparticle2020} or bulk acoustic wave resonators \cite{doelemanBrillouinOptomechanicsQuantum2023}. The experimental approaches used so far to achieve this state start with the system at cryogenic temperatures, in the mK regime inside a dilution refrigerator, decreasing the thermal population of phonons severely. Nonetheless, reaching the quantum ground state has been made possible only by the use of additional laser-based techniques, such as coupling to highly dampened solid state systems \cite{wilson-raeLaserCoolingNanomechanical2004, xueControllableCouplingFlux2007}, feedback cooling \cite{klecknerSubkelvinOpticalCooling2006, arcizetHighSensitivityOpticalMonitoring2006, guoFeedbackCoolingRoom2019, suOptomechanicalFeedbackCooling2023} or resolved sideband cooling \cite{clarkSidebandCoolingQuantum2017, schliesserRadiationPressureCooling2006, marquardtQuantumTheoryCavityAssisted2007, wilson-raeTheoryGroundState2007b, faveroOptomechanicsDeformableOptical2009, giganSelfcoolingMicromirrorRadiation2006}. The latter, also known as dynamical backaction, is based on the engineering of both cavity resonances and laser pump frequencies to favor the cooling over the heating process of a given phonon mode.

Reaching the quantum ground state is often a prerequisite for the study of quantum phenomena and enables applications in precision metrology~\cite{regalMeasuringNanomechanicalMotion2008, cavesMeasurementWeakClassical1980}, phonon thermometry ~\cite{jevticSinglequbitThermometry2015, correaIndividualQuantumProbes2015}, quantum state generation~\cite{bosePreparationNonclassicalStates1997, jahneCavityassistedSqueezingMechanical2009, jaehneGroundstateCoolingNanomechanical2008} or tests of fundamental physics~\cite{marshallQuantumSuperpositionsMirror2003, whittleApproachingMotionalGround2021}. So far, however, the research focus was on optomechanical cavity structures, in which distinct mechanical resonances (standing density waves), interact with specific optical frequencies, thus providing efficient coupling. Despite this fact, a process bound to narrow mechanical resonances limits both the bandwidth of the optomechanical interaction and its application to parallel multi-frequency quantum operations.

An alternative approach is pursued by waveguide optomechanics, in which light can interact with traveling acoustic waves. The optical waves, which can be transmitted at any wavelength in the transparency window of the material, usually interact with a broad continuum of acoustic phonons. This strong interaction is enabled by electrostriction, radiation pressure and photoelasticity, the physical phenomena behind stimulated Brillouin-Mandelstam scattering (SBS) \cite{wolffBrillouinScatteringTheory2021b}. SBS has received great interest for its broad range of applications in optical fibers or integrated photonic waveguides, such as sensing \cite{niklesSimpleDistributedFiber1996, geilenExploringExtremeThermodynamics2022, hotateMeasurementBrillouinGain2000}, signal processing \cite{vidalTunableReconfigurablePhotonic2007, liuIntegratedMicrowavePhotonic2020, zengNonreciprocalVortexIsolator2022}, light storage \cite{merkleinChipintegratedCoherentPhotonicphononic2017, zhuStoredLightOptical2007, stillerCoherentlyRefreshingHypersonic2020}, integrated microwave photonics \cite{marpaungNonlinearIntegratedMicrowave2014, marpaungIntegratedMicrowavePhotonics2019, merkleinStimulatedBrillouinScattering2016} and lasing \cite{hillCwGenerationMultiple1976, otterstromSiliconBrillouinLaser2018, zengOpticalVortexBrillouin2023}. The viable application of SBS for active cooling in waveguides has been theoretically proposed \cite{zhuDynamicBrillouinCooling2023, zhangQuantumCoherentControl2023}. First experimental results show cooling via SBS in optomechanical cavities \cite{bahlObservationSpontaneousBrillouin2012} and integrated silicon waveguides \cite{otterstromOptomechanicalCoolingContinuous2018a}, with cooling rates in the order of tens of Kelvin. Nonetheless, for continuous systems such as waveguides, high SBS cooling rates, particularly high enough to reach the quantum ground state, are still an open challenge.

Here, we experimentally demonstrate optoacoustic cooling of a band of continuous traveling acoustic waves in a waveguide system at room temperature. Without using a cryogenic environment, the acoustic phonons at 7.38\,GHz are cooled by 219\,K, reaching an effective mode temperature of 74\,K. The cooled acoustic waves extend over a macroscopic length of 50\,cm in a tapered chalcogenide glass photonic crystal fiber (PCF). The asymmetry of anti-Stokes and Stokes processes in the considered backward SBS interaction provides natural symmetry breaking in the heating-cooling of phonons and therefore no sideband cooling is necessary. We underpin our experimental results with a theoretical model that reproduces the replenishing of acoustic phonons by dissipation in a strong Brillouin-Mandelstam cooling regime. Given the long interaction length of 50\,cm, the mechanical object addressed in the interaction is more massive than standard micro resonators. Achieving ground state cooling of such a macroscopic phonon would pave the way towards exploring the transition of classical to quantum physics.
\begin{figure}
\includegraphics{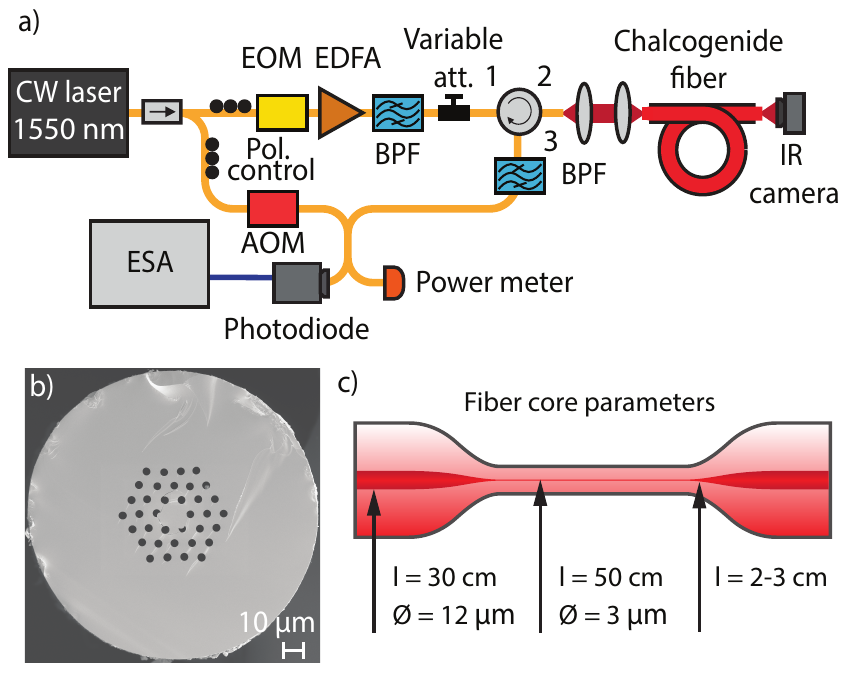}
\caption{(Color online) (a) Diagram of the experimental setup used for the measurement of the backscattered SBS signal as a function of input power via heterodyne detection. CW: continuous wave, EOM (AOM): electro-optical (acousto-optical) modulator, EDFA: erbium-doped fiber amplifier, BPF: band-pass filter, ESA: electrical spectrum analyzer. (b) Scanning electron microscope image of the cross-section of the untapered chacogenide glass PCF. The air-filling ratio of this fiber is 0.48 and it has a core diameter of 12\,$\upmu$m. (c) Diagram of the tapered sample. The taper waist has a length (l) of 50\,cm and a core diameter of 3\,$\upmu$m. The waist is connected to 30\,cm sections of untapered  fiber, one on each side, via transitions of 2-3\,cm in length. }
\label{fig1}
\end{figure}
\section{Experimental setup}
The setup used for the experiment is shown in Fig.\,\ref{fig1}(a). The output of a continuous wave (CW) laser at wavelength $\lambda_p=$\,1550\,nm is divided in two branches, pump and local oscillator (LO). The pump light is modulated into 100\,ns long square pulses, with 25\,\% duty cycle, amplified via an erbium-doped fiber amplifier (EDFA) and filtered with a band-pass filter (BPF). The fixed output of the EDFA can be controlled via a variable attenuator, allowing to study the SBS resonances as a function of pump power. The coupling into the sample, a tapered chalcogenide glass solid core single mode PCF from Selenoptics, is done via free space. The glass composition is $\mathrm{Ge_{10}As_{22}Se_{68}}$, with a SBS resonance of $\Omega_B/2\pi$\,(1550 nm)\,=\,7.38\,GHz. The PCF air-filling ratio is 0.48 (Fig.\,\ref{fig1}(b)) and the insertion and transmission loss through the fiber is -3.64\,dB in total. The initial core diameter is 12\,$\upmu$m, which is tapered down to 3\,$\upmu$m at the waist, with a length of 50 cm (Fig.\,\ref{fig1}(c)). An infrared (IR) camera allows to visualize the optical mode propagating through the core and optimize coupling. As the SBS signal is backscattered, a circulator stops it from going back into the laser, redirecting it into the detection part of the setup. Given the high Fresnel coefficients of the fiber facets, another BPF is used to filter out the strong elastic pump back-reflection. The filtered signal is mixed with a frequency-shifted LO, via a 200\,MHz acousto-optic modulator (AOM), to perform heterodyne detection. The resulting optical interference is detected with a photodiode and the transduced electrical signal measured with an electrical spectrum analyzer (ESA). The experiment is performed at room temperature (293\,K).
\begin{figure}
\includegraphics{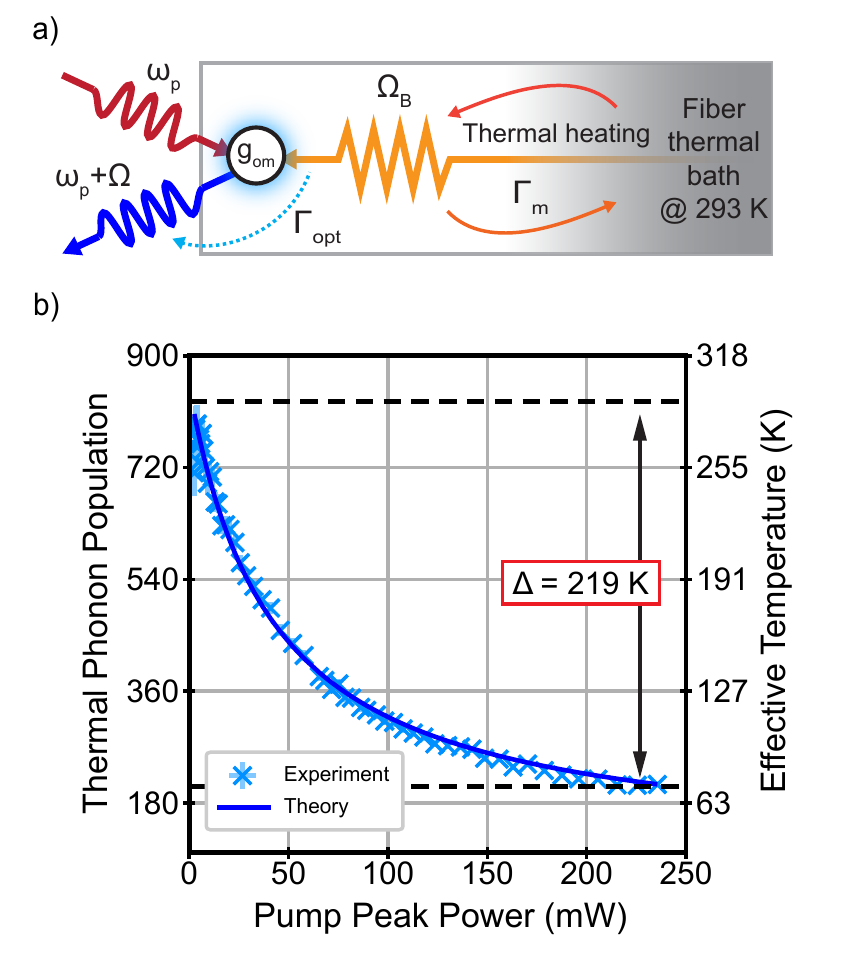}
\caption{(Color online) (a) Diagram of the different processes affecting the population of resonant anti-Stokes phonons at $\Omega_B$. (b) Light blue crosses show the experimentally measured decrease of resonant anti-Stokes phonon population ($\mathrm{\Omega_B}/2\pi$\,=\,7.38\,GHz) and its respective effective temperature as a function of pump power. From an initial population at room temperature (293\,K) of 830 phonons, a final population of 212 is measured, corresponding to 74\,K. This results in a temperature decrease of 219\,K, or 74.7\,\%. The solid blue line shows the theoretical decrease of phonon population according to Eq.\,(\ref{phonon occupation steady state}). The horizontal black dashed lines are added as visual aids and indicate the initial and final temperature of the phonon mode.}
\label{fig2}
\end{figure}
\section{Experimental results}
A diagram of the different mechanisms affecting the population of SBS-resonant anti-Stokes phonons $\Omega_B$ is shown in Fig.\,\ref{fig2}(a). In this type of scattering, phonons are annihilated. Therefore, the interaction can be understood as a loss mechanism, present only while the system is optically pumped. The other two processes that affect the phonon occupation are thermal heating and acoustic dissipation. The population of a phonon mode ($\mathrm{n_{th}}$) at a given frequency ($\Omega$) is given by the Bose-Einstein statistics $\mathrm{n_{th}(\Omega)=1/{(e}^{\hbar \Omega/k_BT}-1)}$ and will depend on the temperature of the system (T). As the system thermalises,  the phonon levels are filled with incoherent phonons. Acoustic dissipation, on the other hand, describes the decay of the traveling density fluctuation. While a SBS interaction takes place, energy is being transferred from the acoustic into the optical field, with a coupling strength $g_{om}$. This results in an effective temperature decrease of the resonant phonon mode.

The shape of the SBS resonances is described by $\mathrm{G_B(\omega)=g_B(\Gamma_{eff}/2)^2/((\Omega_B-\omega)^2+(\Gamma_{eff}/2)^2})$ where $\mathrm{g_B}$ is the intrinsic nonlinear gain of the sample and the effective dissipation rate ($\mathrm{\Gamma_{eff}}$) is given by the full width at half maximum (FWHM). The effective dissipation rate is defined as $\mathrm{\Gamma_{eff}=\Gamma_m+\Gamma_{opt}}$, where $\mathrm{\Gamma_m}$ is the natural acoustic dissipation rate and $\mathrm{\Gamma_{opt}}$ is the optically induced loss resulting from the SBS interaction. The cooling rate (R) is defined as the ratio between final and initial phonon occupation, $\bar{n}_f$ and $\bar{n}_0$ respectively. In the weak coupling regime R is given by

 \begin{equation}
    R= \mathrm{\frac{\bar{n}_f}{\bar{n}_0}=\frac{\Gamma_m}{\Gamma_{eff}}}\,.
 \end{equation}
From the final phonon population, the effective temperature of the mode after the active cooling can be obtained using the Bose-Einstein equation. In Fig.\,\ref{fig2}(b) the experimental results for the tapered chalcogenide PCF are shown. From an initial phonon population of 830 at 293\,K, a final population of 212 phonons is measured. This corresponds to an effective temperature of 74\,K, resulting in a decrease of 219\,K or 74.7\,\% from room temperature. Most of the SBS response from the sample comes from the waist of the taper, which is 50\,cm long. The phonons addressed in the nonlinear interaction extend all over the active part, resulting thus in a phonon of macroscopic mass being cooled down to cryogenic temperatures.
\begin{figure}
\includegraphics{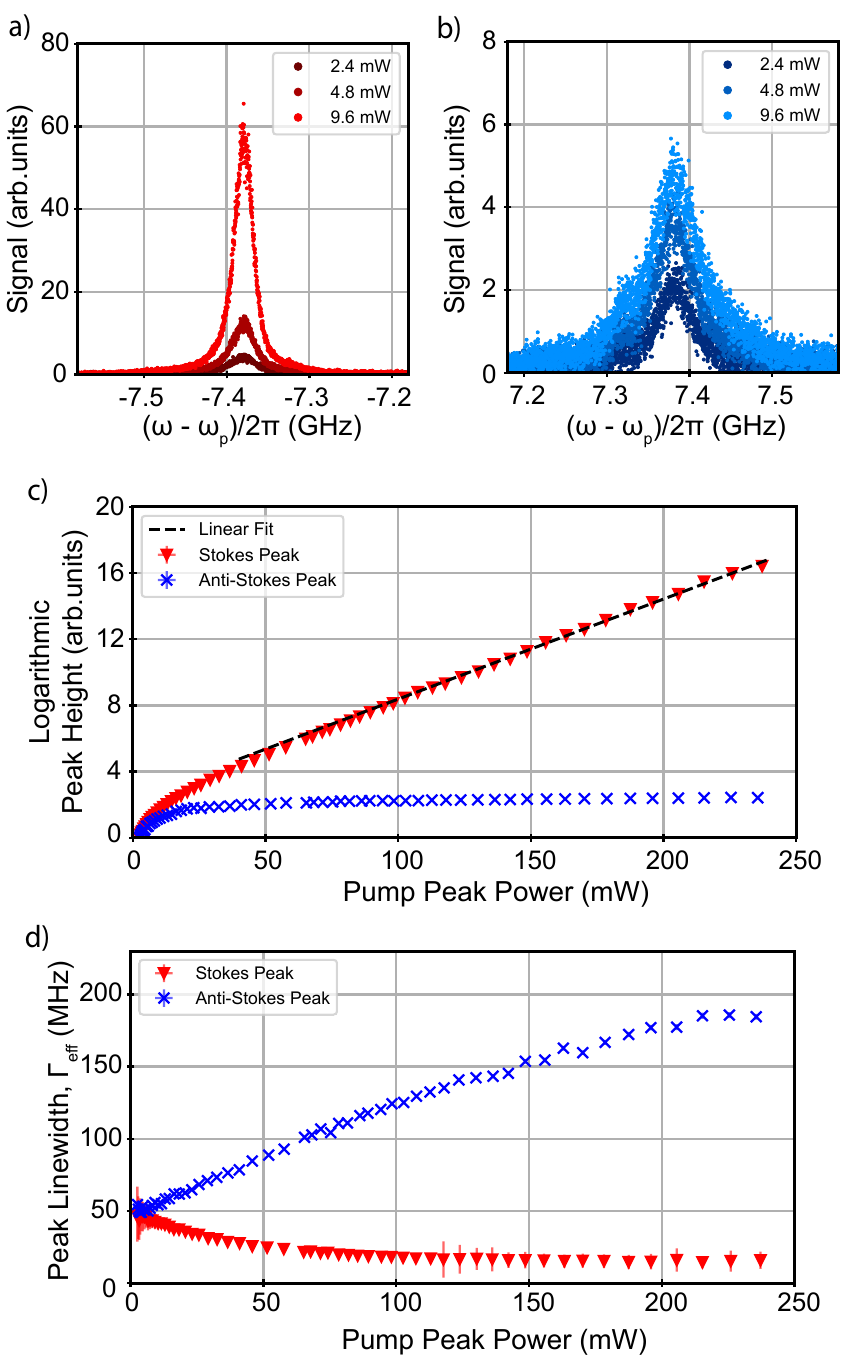}
\caption{(Color online) (a) In red, behaviour of the Stokes resonance for three different pump powers. (b) In blue, anti-Stokes resonance for the same powers as (a). Note the different y-axis scales.  (c) Peak height of the Stokes (red triangles) and anti-Stokes (blue crosses) resonances in logarithmic scale as a function of pump power. (d) Peak linewidth ($\mathrm{\Gamma_{eff}}$) of the Stokes (red triangles) and anti-Stokes (blue crosses) resonances as a function of pump power. The Stokes resonance narrows, but the anti-Stokes peaks broadens. This indicates an increase of the effective dissipation rate ($\mathrm{\Gamma_{eff}=\Gamma_m+\Gamma_{opt}}$) of the addressed phonons, caused by the SBS interaction and a relaxation of the phase-matching condition.}
\label{fig3}
\end{figure}
In an optomechanical resonator, both the Stokes and anti-Stokes processes address the same phonon field. For backward SBS in a waveguide system, such as an optical fiber, this symmetry is broken. The resonant longitudinal phonons involved in a Stokes process are co-propagating with the pump, while for the anti-Stokes process, they are counter-propagating. This inherent symmetry breaking allows to perform the cooling shown without working in the resolved sideband regime. Additionally, both resonances can be studied simultaneously, allowing to compare their different behaviours (Fig.\,\ref{fig3}(a) and \ref{fig3}(b)). The SBS peaks, with a Lorentzian shape, are defined by two parameters, height and linewidth ($\mathrm{\Gamma_{eff}}$).

The evolution of the peak height as a function of pump power is shown in Fig.\,\ref{fig3}(c). For low pump powers, both resonances increase in a parallel way, as the initial equilibrium population of the addressed phonon baths are almost equal. After a threshold of 15\,mW, the Stokes peak increases exponentially. This behaviour is characteristic of stimulated scattering, in which the interference between pump and scattered light and the acoustic waves drive each other, creating a feedback loop. The linear fit of the exponential increase provides the SBS gain of the sample, g$_B$\,=\,(1.32\,±\,0.18)\,·\,10$^{-9}$\,m/W, comparable with literature values \cite{abedinObservationStrongStimulated2005a, ogusuBrillouingainCoefficientsChalcogenide2004}. The anti-Stokes peak, on the contrary, saturates in height after this threshold. Regarding the linewidths, different behaviours are observed (Fig.\,\ref{fig3}(d)). The Stokes resonance narrows, as expected for SBS, while the anti-Stokes broadens. Both the broadening and saturation are footprints of the cooling of phonons via SBS. In the case of a Stokes interaction, the energy is transferred from the optical to the acoustic field, and the amount of phonons that can be created is limited by pump depletion or the damage threshold of the sample. Therefore, an increase in peak height is observed. This is not the case for the anti-Stokes resonance. The number of scattered photons depends on the available number of phonons, fixed by the initial temperature of the thermal bath. As the pump power is increased, more phonons are actively removed from the system, but a limit will be reached. In the ideal case, an observation of peak height saturation would mean that the system has entered a new equilibrium state, in which the thermal bath replenishes the mode at the same speed at which the phonons are actively annihilated. In our experiment this is not the case, as pump depletion arising from the strong Stokes interaction was observed to limit the cooling power \cite{[{See Supplemental Material at }][{ for additional derivations of the equation of linearized SBS interaction, analysis of Brillouin-Mandelstam cooling via the covariance approach and an experimental study of the pump depletion via the Stokes wave limiting cooling.}]supp}.
The linewidth broadening describes how broad the resonance condition is, i.e., how far from perfect phase-matching the process can still occur. As the pump is increased, more perfectly phase-matched phonons are removed, yet more photons are present. The probability of scattering with an off-resonance phonon therefore broadens.

\section{Theory of Brillouin cooling in waveguides} \label{theory}
An analysis using the theory of waveguide Brillouin optomechanics~\cite{zhangQuantumCoherentControl2023,vanlaerUnifyingBrillouinScattering2016,rakichQuantumTheoryContinuum2018} of the experimental results is presented in this section. In a typical SBS-active waveguide, backward SBS describes an optoacoustic interaction where two light fields are coherently coupled to an acoustic field. By absorbing a pump photon, the frequency of the backward-scattered photons can be upshifted or downshifted, which corresponds to the Stokes and anti-Stokes processes, respectively. The Stokes process is a parametric down-conversion interaction. It causes heating for acoustic phonons and enables the generation of entangled photon-phonon pairs. The anti-Stokes process is a beam-splitter interaction between scattered photons and acoustic phonons. It can produce phonon cooling, i.e., broadening of the acoustic linewidth, as shown in Figs.\,\ref{fig3}(b) and \ref{fig3}(d). In addition, the natural dispersive symmetry breaking between the Stokes and anti-Stokes processes in the backward SBS scattering in waveguides allows to study the anti-Stokes process individually.
By giving the system Hamiltonian derived previously in~\cite{sipeHamiltonianTreatmentStimulated2016,zoubiOptomechanicalMultimodeHamiltonian2016} and considering the undepleted pump approximation~\cite{chenBrillouinCoolingLinear2016a}, the dynamics of the linearized optoacoustic anti-Stokes interaction in the momentum space can be given by \cite{supp}
\begin{eqnarray}\label{Dynamic equation of linearized case}
\frac{d a_{as}}{dt} &=& \left[ -i(-\Delta_L + \Delta_1) - \frac{\gamma_{\rm o}}{2}  \right] a_{as}
- i g_{\rm om} b_{ac} + \sqrt{\gamma_{\rm o}}\xi_{as}, \nonumber\\
\frac{d b_{ac}}{dt} &=& \left[ -i( \Omega_{\rm B} + \Delta_2 ) - \frac{\Gamma_{\rm m}}{2} \right] b_{ac} - i g_{\rm om} a_{as} + \sqrt{\Gamma_{\rm m}}\xi_{ac}, \nonumber\\
\end{eqnarray}
where $a_{as}$ ($b_{ac}$) denote the photon (phonon) annihilation operator for the $k^{th}$ anti-Stokes 
mode (acoustic mode) with wavenumber $k$. $\Delta_L$ is the frequency detuning between  pump and anti-Stokes fields. $\gamma_{\rm o}$ and $g_{\rm om}$ correspond to the optical loss rate and the pump-enhanced optoacoustic coupling strength between anti-Stokes photons and acoustic phonons. $\Delta_1 = k\upsilon_{\rm as}$ and $\Delta_2=k\upsilon_{\rm ac}$ represent wavenumber-induced frequency shifts for the anti-Stokes photons and acoustic phonons, where $\upsilon_{\rm as}$ ($\upsilon_{\rm ac}$) is the group velocity of the anti-Stokes (acoustic) wave. $\xi_{as}$ denotes the quantum zero-mean Gaussian noise of the anti-Stokes mode and $\xi_{ac}$ corresponds to the acoustic thermal noise which obeys relations $\langle \xi_{ac}(t) \rangle = 0$ and $\langle \xi_{ac}^{\dagger}(t_1) \xi_{ac}(t_2)\rangle=n_{\rm th}\delta(t_1-t_2)$, where $n_{\rm th}$ is the thermal phonon occupation under the environment temperature.

For simplicity, only the case where the anti-Stokes mode and acoustic mode are phase-matched with pump mode, i.e., $\Delta_1=\Delta_2=0$ and $\Delta_L = -\Omega_{\rm B}$ is discussed.
By switching to a frame rotating with frequency $\Omega_{\rm B}$ and considering relations of Langevin noises $\xi_{as,ac}$, the dynamics of the mean phonon number and photon number can be given by~\cite{zhuDynamicBrillouinCooling2023}
\begin{eqnarray}\label{Dynamics of mean photon or phonon number}
\dot{N}_a &=& -\gamma_{\rm o} N_a - i g_{\rm om}( \langle a_{as}^{\dagger}b_{ac}\rangle - \langle a_{as}^{\dagger}b_{ac}\rangle^{*} ), \nonumber\\
\dot{N}_b &=& -\Gamma_{\rm m} N_b + i g_{\rm om}( \langle a_{as}^{\dagger}b_{ac}\rangle - \langle a_{as}^{\dagger}b_{ac}\rangle^{*} ) + \Gamma_{\rm m}n_{\rm th}, \nonumber\\
\dot{\langle a_{as}^{\dagger}b_{ac}\rangle} &=& - \frac{\gamma_{\rm o}+\Gamma_{\rm m}}{2}\langle a_{as}^{\dagger}b_{ac}\rangle - i g_{\rm om}N_a + i g_{\rm om}N_b,
\end{eqnarray}
where ${N}_a=\langle a_{as}^{\dagger}a_{as}\rangle$ and ${N}_b=\langle a_{ac}^{\dagger}a_{ac}\rangle$ correspond to the mean photon and phonon numbers, respectively. Eq.~(\ref{Dynamics of mean photon or phonon number}) can be solved to obtain the phonon occupation at the steady state \cite{supp}
\begin{eqnarray}\label{phonon occupation steady state}
N_{b}^{\rm ss} = \frac{4g_{\rm om}^2 + \gamma_{\rm o}(\gamma_{\rm o}+\Gamma_{\rm m})}{ 4g_{\rm om}^2 + \gamma_{\rm o}\Gamma_{\rm m} } \cdot \frac{\Gamma_{\rm m}}{\gamma_{\rm o}+\Gamma_{\rm m}}n_{\rm th}.
\end{eqnarray}
From Eq.~(\ref{phonon occupation steady state}), the cooling rate can be enhanced by increasing the coupling strength $g_{\rm om}$, i.e., increasing the pump power, as shown in Fig.~\ref{fig2} (b). However, this cooling rate will be limited by the ratio $\Gamma_{\rm m}/(\gamma_{\rm o}+\Gamma_{\rm m})$, similar to the case of sideband cooling in cavity optomechanics~\cite{aspelmeyerCavityOptomechanics2014,genesGroundstateCoolingMicromechanical2008,qiuLaserCoolingNanomechanical2020}. 
The optically-enhanced acoustic damping rate is defined thus as
\begin{eqnarray}\label{Gamma effective}
\Gamma_{\rm eff} = \Gamma_{\rm m} + \frac{4g_{\rm om}^2\gamma_{\rm o}}{ 4g_{\rm om}^2 + \gamma_{\rm o}(\Gamma_{\rm m}+\gamma_{\rm o}) },
\end{eqnarray}
which can be seen in Figs.\,\ref{fig3}(b) and \ref{fig3}(d), as the anti-Stokes resonance broades with increasing input power. Eq.\,(\ref{Gamma effective}) shows a saturation in linewidth for high pump powers, indicating a physical limit for the phonon cooling achievable through this process. Given the system parameters in this experiment, the minimum phonon population achievable from room temperature is around 100 phonons, corresponding to an effective temperature of 36\,K ($R=0.1$). It should be noted that this system is a continuous optomechanical system, which provides cooling for groups of phonons~\cite{otterstromOptomechanicalCoolingContinuous2018a,zhuDynamicBrillouinCooling2023}, instead of single-mode or multi-mode mechanical cooling in cavity optomechanical systems. In Eq.~(\ref{Dynamics of mean photon or phonon number}), only  the phase-matching case with zero wavenumber is considered. For acoustic modes with non-zero wavenumber, the cooling rate at steady-state can be calculated by including the effects of the wavenumber-induced frequency shifts $\Delta_{1,2}$ \cite{supp}. 

\section{Conclusions and outlook}
This experiment has demonstrated that SBS is a promising tool in the challenge of bringing waveguide modes to their quantum ground state of mechanical motion. A massive 50\,cm-long phonon with frequency 7.38\,GHz is brought to cryogenic temperatures from room temperature, reducing the effective mode temperature by 219\,K, one order of magnitude higher than previously reported \cite{otterstromOptomechanicalCoolingContinuous2018a}. With our novel theoretical description of the cooling process, the physical cooling limit was calculated to be 90\,\% of population decrease. This opens the path to the realistic achievement of reaching the quantum ground state in waveguides, given the high frequency of the resonant phonons addressed in this experiment. Performing the experiment in a cryogenic environment, such as a liquid helium cryostat at 4\,K, paired with the efficient cooling present in the fiber,  would produce occupations of few or even less than one phonon. These results therefore pave the way towards accessing the quantum nature of massive objects. A similar work was published on the arXiv \cite{johnsonLaserCoolingTraveling2023} showing cooling by 21\,K in a liquid-core fiber.
\begin{acknowledgments}
We thank C. Silberhorn and C. Wolff and our co-workers A. Popp, X. Zeng, S. Becker, Z. O. Saffer, and J. Landgraf for valuable discussions. We acknowledge funding from the Max Planck Society through the Independent Max Planck Research Group scheme.
\end{acknowledgments}

%

\clearpage

\newpage

\appendix
\onecolumngrid

\maketitle This Supplementary Material provides additional analyses and
derivations on the following topics: (\ref{S1}) motion equation of linearized Brillouin
interaction, (\ref{S2}) analyzing Brillouin cooling via the covariance approach and (\ref{S3}) pump depletion limiting cooling.

\section{Motion equation of linearized backward Brillouin anti-Stokes interaction}\label{S1}
For backward Brillouin anti-Stokes scattering, based on the Hamiltonian derived in~\cite{sipeHamiltonianTreatmentStimulated2016_supp, zoubiOptomechanicalMultimodeHamiltonian2016_supp}, the quantum free Hamiltonian of the pump field, anti-Stokes field, and acoustic field is given by
\begin{eqnarray}
H_{p}^{\rm free} &=& \int \hbar \omega_p(k) a_p^{\dagger}(k)a_p(k)dk, \nonumber\\
H_{as}^{\rm free} &=& \int \hbar \omega_{as}(k) a_{as}^{\dagger}(k)a_{as}(k)dk, \nonumber\\
H_{ac}^{\rm free} &=& \int \hbar \omega_{ac}(q) b_{ac}^{\dagger}(q)b_{ac}(q)dq,
\end{eqnarray}
and the interaction Hamiltonian can be expressed as
\begin{eqnarray}
H_{\rm int} = \hbar g_0 \int \int a_{as}^{\dagger}(k) a_p(k-q)b_{ac}(q)dkdq + h.c., 
\end{eqnarray}
where $a_p$, $a_{as}$, and $b_{ac}$ denote annihilation operators for the $k^{th}$ pump mode, anti-Stokes mode, and acoustic mode with wavenumber ($k$ or $q$), respectively. $\omega_p$, $\omega_{as}$, and $\omega_{ac}$ correspond to the carrier frequencies of these three modes. $g_0$ is the vacuum coupling strength~\cite{vanlaerUnifyingBrillouinScattering2016_supp}.
We define the pump mode as follows
\begin{eqnarray}
a_p(k_p) = \frac{1}{\sqrt{L}}\int_0^{L} A_p(z) e^{-i(k_p-\Delta k)z} dz,
\end{eqnarray}
where $\Delta k = k-q$. $A_p$ is the envelope operator of the pump field and $L$ is the length of the waveguide. Considering the undepleted pump approximation, the three-wave interaction can be reduced to an effective interaction between the anti-Stokes mode and acoustic mode with a linearized interaction Hamiltonian
\begin{eqnarray}
H_{\rm int}^{\rm lin} = \hbar g_{\rm om} \int \int a_{as}^{\dagger}(k)b_{ac}(q)dkdq + h.c.,
\end{eqnarray}
where $g_{\rm om} = g_0 A_p \sqrt{L}$ is the pump-enhanced effective optomechanical coupling strength. Thus the total Hamiltonian of the linearized Brillouin anti-Stokes scattering can be written as
\begin{eqnarray}
H = \int \hbar \omega_{as}(k) a_{as}^{\dagger}(k)a_{as}(k)dk + 
\int \hbar \omega_{ac}(q) b_{ac}^{\dagger}(q)b_{ac}(q)dq
+  \hbar g_{\rm om} \int \int a_{as}^{\dagger}(k)b_{ac}(q)dkdq + h.c..
\end{eqnarray}
Applying $H$ to the Heisenberg equation, the dynamics of the linearized anti-Stokes process can be given by
\begin{eqnarray}
\frac{d a_{as}}{dt} &=& ( -i\omega_{as}(k) - \frac{\gamma_{\rm o}}{2} ) a_{as}(k)
- i g b_{ac}(q) + \sqrt{\gamma_{\rm o}} \xi_{as}, \nonumber\\
\frac{d b_{ac}}{dt} &=& ( -i\omega_{ac}(q) - \frac{\Gamma_{\rm m}}{2} ) b_{ac}(q)
- i g a_{as}(k) + \sqrt{\Gamma_{\rm m}}\xi_{ac},
\end{eqnarray}
where $\gamma_{\rm o}$ and $\Gamma_{\rm m}$ denote the loss rates of the anti-Stokes and acoustic modes, respectively. $\xi_{as}$ and $\xi_{ac}$ are the quantum Langevin noises which obey relations~\cite{chenBrillouinCoolingLinear2016a_supp, zhuDynamicBrillouinCooling2023_supp}
\begin{eqnarray}
\langle \xi_{as}(t,k) \rangle &=& 0, \nonumber\\
\langle \xi_{ac}(t,q) \rangle &=& 0, \nonumber\\
\langle \xi_{as}^{\dagger}(t_1,k)\xi_{as}(t_2,k)\rangle &=& 0, \nonumber\\
\langle \xi_{ac}^{\dagger}(t_1,q)\xi_{ac}(t_2,q)\rangle &=& n_{th}\delta(t_1-t_2),
\end{eqnarray}
where $n_{th}=1/(e^{\hbar\omega_{ac}/k_B T_{\rm en}}-1)$ is the thermal phonon occupation
at the environment temperature $T_{\rm en}$. In addition, the anti-Stokes photon and acoustic phonon dispersion relations with the first-order approximation can be expressed as~\cite{sipeHamiltonianTreatmentStimulated2016_supp, kharelNoiseDynamicsForward2016_supp}
\begin{eqnarray}
\omega_{as}(k) &=& \omega_{as,0} + \upsilon_{\rm o} ( k - k_{as,0} ), \nonumber\\
\omega_{ac}(k) &=& \omega_{ac,0} + \upsilon_{\rm ac} ( q - q_{ac,0} ),
\end{eqnarray}
where $\omega_{as,0}$ ($\omega_{ac,0}$) and $k_{as,0}$ ($q_{ac,0}$) represent the central frequency and carrier wave vector of the scattered anti-Stokes field (acoustic field), respectively. $\upsilon_{\rm o}$ and $\upsilon_{\rm ac}$ are the group velocities of the anti-Stokes and acoustic fields, Thus motion equations of $a_{as}$ and $b_{ac}$ can be rewritten as
\begin{eqnarray}
\frac{d a_{as}}{dt} &=& \left[ -i(\omega_{as,0}+\Delta_1) - \frac{\gamma_{\rm o}}{2} \right] a_{as}(k) - i g_{om} b_{ac}(q) + \sqrt{\gamma_{\rm o}} \xi_{as}, \nonumber\\
\frac{d b_{ac}}{dt} &=& \left[ -i(\omega_{ac,0}+\Delta_2) - \frac{\Gamma_{\rm m}}{2} \right] b_{ac}(q) - i g_{om} a_{as}(k) + \sqrt{\Gamma_{\rm m}}\xi_{ac},
\end{eqnarray}
where $\Delta_1=k\upsilon_{\rm o}$ and $\Delta_2=q\upsilon_{\rm ac}$ imply the wavenumber-induced frequency shifts for the $k$th anti-Stokes mode and $q$th acoustic mode. $\Delta_1=\Delta_2=0$ corresponds to the case that the anti-Stokes and acoustic modes are phase-matched with the pump mode.

Switching to a frame rotating at the pump frequency $\omega_{p,0}$, we have
\begin{eqnarray}
\frac{d a_{as}}{dt} &=& \left[ -i(-\Delta_L+\Delta_1) - \frac{\gamma_{\rm o}}{2} \right] a_{as}(k) - i g_{om} b_{ac}(q) + \sqrt{\gamma_{\rm o}} \xi_{as}, \nonumber\\
\frac{d b_{ac}}{dt} &=& \left[ -i(\omega_{ac,0}+\Delta_2) - \frac{\Gamma_{\rm m}}{2} \right] b_{ac}(q) - i g_{om} a_{as}(k) + \sqrt{\Gamma_{\rm m}}\xi_{ac},
\end{eqnarray}
where $\Delta_L = \omega_{p,0} - \omega_{as,0}$ is the frequency detuning between the pump mode and the anti-Stokes mode. Under the phase-matching condition of the anti-Stokes Brillouin scattering, i.e., $\omega_{p,0} - \omega_{as,0} = -\omega_{ac,0} = -\Omega_{\rm B}$, the motion equations of $a_{as}$ and $b_{ac}$ can be re-expressed as
\begin{eqnarray}
\frac{d a_{as}}{dt} &=& \left[ -i(\Omega_{\rm B}+\Delta_1) - \frac{\gamma_{\rm o}}{2} \right] a_{as}(k) - i g_{om} b_{ac}(q) + \sqrt{\gamma_{\rm o}} \xi_{as}, \nonumber\\
\frac{d b_{ac}}{dt} &=& \left[ -i(\Omega_{\rm B}+\Delta_2) - \frac{\Gamma_{\rm m}}{2} \right] b_{ac}(q) - i g_{om} a_{as}(k) + \sqrt{\Gamma_{\rm m}}\xi_{ac},
\end{eqnarray}
where $\Omega_{\rm B}$ is the Brillouin frequency. We switch to a frame rotating at the Brillouin frequency $\Omega_{\rm B}$ for simplification and thus reduce the dynamics of the linearized anti-Stokes process as follows
\begin{eqnarray}\label{Dynamics of linearized anti-Stokes process}
\frac{d a_{as}}{dt} &=& (-i\Delta_1 - \frac{\gamma_{\rm o}}{2} ) a_{as}(k) - i g_{om} b_{ac}(q) + \sqrt{\gamma_{\rm o}} \xi_{as}, \nonumber\\
\frac{d b_{ac}}{dt} &=& ( -i\Delta_2 - \frac{\Gamma_{\rm m}}{2} ) b_{ac}(q) - i g_{om} a_{as}(k) + \sqrt{\Gamma_{\rm m}}\xi_{ac}.
\end{eqnarray}

\section{Analyzing Brillouin cooling via the covariance approach }\label{S2}
In the linear regime under the undepleted pump approximation, the mean phonon number can be exactly calculated by solving a linear system of differential equations involving the second-order moments of the anti-Stokes and acoustic modes~\cite{zhuDynamicBrillouinCooling2023_supp, liuReviewCavityOptomechanical2013_supp}.
The motion equations of the mean photon number $N_a = \langle a_{as}^{\dagger}a_{as}\rangle$ and the mean phonon number $N_b =\langle b_{ac}^{\dagger}b_{ac}\rangle$ can be derived for Eq.~(\ref{Dynamics of linearized anti-Stokes process}) which have been derived previously in our recent work~\cite{zhuDynamicBrillouinCooling2023_supp} in the strong coupling regime. Now we derive the motion equations of $N_{a,b}$ in the weak coupling regime for the case of our experiment, which is similar to the method used in~\cite{zhuDynamicBrillouinCooling2023_supp}. Based on Eq.~(\ref{Dynamics of linearized anti-Stokes process}), we have
\begin{eqnarray}\label{Motion equation of second order moments with noise terms1}
\frac{d N_a}{d t} &=& -\gamma_o N_a - i g_{om} ( \langle a_{as}^{\dagger} b_{ac}\rangle - \langle a_{as}^{\dagger} b_{ac} \rangle^* )
+ \sqrt{\gamma_o}( \langle \xi_{as}^{\dagger} a_{as}\rangle + \langle \xi_{as}^{\dagger} a_{as}\rangle^* ), \nonumber\\
\frac{d N_b}{d t} &=& -\Gamma_m N_b + i g_{om} ( \langle a_{as}^{\dagger} b_{ac}\rangle - \langle a_{as}^{\dagger} b_{ac} \rangle^* )
+ \sqrt{\Gamma_m} ( \langle \xi^{\dagger}_{ac} b_{ac} \rangle + \langle \xi^{\dagger}_{ac} b_{ac} \rangle^* ), \nonumber\\
\frac{d \langle a_{as}^{\dagger}b_{ac}\rangle}{d t} &=& - \left (i(\Delta_1-\Delta_2)+\frac{\gamma_o+\Gamma_m}{2}\right )
\langle a_{as}^{\dagger}b_{ac}\rangle - i g_{om} N_a + i g_{om} N_b + \sqrt{\gamma_o}\langle \xi_{as}^{\dagger} b_{ac} \rangle
+ \sqrt{\Gamma_m} \langle a_{as}^{\dagger} \xi_{ac}\rangle.
\end{eqnarray}

Now we calculate the noise-related terms in Eq.~(\ref{Motion equation of second order moments with noise terms1}). In fact, If we apply the Fourier transform to Eq.~(\ref{Dynamics of linearized anti-Stokes process}), the analytical solutions can be expressed as
\begin{eqnarray}\label{Analitical solution of photon and phonon in frequency domain}
\bar{a}_{as}(\omega) &=& - \frac{ \left[ i(\omega+\Delta_2)-\frac{\Gamma_{\rm m}}{2} \right]\sqrt{\gamma_{\rm o}}\bar{\xi}_{as} + i g_{\rm om}\sqrt{\Gamma_{\rm m}} \bar{\xi}_{ac} }
{ g_{\rm om}^2 + \frac{\gamma_{\rm o}\Gamma_{\rm m}}{4} - (\omega+\Delta_1)(\omega+\Delta_2) - i\left[ \frac{\gamma_{\rm o}}{2}(\omega+\Delta_2) + \frac{\Gamma_{\rm m}}{2}(\omega+\Delta_1) \right] }, \nonumber\\
\bar{b}_{ac}(\omega) &=& - \frac{ i g_{\rm om}\sqrt{\gamma_{\rm o}}\bar{\xi}_{as} + \left[ i(\omega+\Delta_1)-\frac{\gamma_{\rm o}}{2}\right]\sqrt{\Gamma_{\rm m}}\bar{\xi}_{ac} }{ g_{\rm om}^2 + \frac{\gamma_{\rm o}\Gamma_{\rm m}}{4} - (\omega+\Delta_1)(\omega+\Delta_2) - i\left[ \frac{\gamma_{\rm o}}{2}(\omega+\Delta_2) + \frac{\Gamma_{\rm m}}{2}(\omega+\Delta_1) \right] },
\end{eqnarray}
where $\bar{a}_{as}(\omega)$ and $\bar{b}_{ac}(\omega)$ are the Fourier transform of $a_{as}$ and $b_{ac}$, which obey
\begin{eqnarray}
a_{as}(t) &=& \frac{1}{2\pi} \int_{-\infty}^{\infty} \bar{a}_{as}(\omega)e^{-i\omega t} d\omega, \nonumber\\
b_{ac}(t) &=& \frac{1}{2\pi} \int_{-\infty}^{\infty} \bar{b}_{ac}(\omega)e^{-i\omega t} d\omega. 
\end{eqnarray}
$\bar{\xi}_{as}$ and $\bar{\xi}_{ac}$ are the Fourier transform of Langevin noises $\xi_{as}$ and $\xi_{ac}$, respectively, which obeys relations as follows
\begin{eqnarray}\label{Noise relations in frequency domain}
\langle \bar{\xi}_{as}^{\dagger}(\omega_1)\bar{\xi}_{as}(\omega_2) \rangle &=& 0, \nonumber\\
\langle \bar{\xi}_{ac}^{\dagger}(\omega_1)\bar{\xi}_{ac}(\omega_2) \rangle &=& 2\pi n_{th}\delta(\omega_1-\omega_2), \nonumber\\
\langle \bar{\xi}_{as}^{\dagger}(\omega_1)\bar{\xi}_{ac}(\omega_2) \rangle &=& 0.
\end{eqnarray}
Now we can calculate the noise-related terms in Eq.~(\ref{Motion equation of second order moments with noise terms1}) by using Eqs.~(\ref{Analitical solution of photon and phonon in frequency domain})-(\ref{Noise relations in frequency domain}) as follows
\begin{eqnarray}
&&\langle \xi_{ac}^{\dagger}(t) a_{as}(t) \rangle \nonumber \\
&& = \frac{1}{4\pi^2} \int_{-\infty}^{\infty} e^{i\omega_1 t} d\omega_1 
\int_{-\infty}^{\infty} e^{-i\omega_2 t} \langle \bar{\xi}_{as}^{\dagger}(\omega_1) \bar{a}_{as}(\omega_2) \rangle d\omega_2 \nonumber\\
&& = -\frac{1}{4\pi^2} \int_{-\infty}^{\infty} e^{i\omega_1 t} d\omega_1 
\int_{-\infty}^{\infty} e^{-i\omega_2 t}
\frac{ \left[ i(\omega_2+\Delta_2)-\frac{\Gamma_{\rm m}}{2} \right]\sqrt{\gamma_{\rm o}} 
\langle \bar{\xi}_{as}^{\dagger}(\omega_1) \bar{\xi}_{as}(\omega_2) \rangle + i g_{\rm om}\sqrt{\Gamma_{\rm m}} \langle \bar{\xi}_{as}^{\dagger}(\omega_1) \bar{\xi}_{ac}(\omega_2) \rangle  }{ g_{\rm om}^2 + \frac{\gamma_{\rm o}\Gamma_{\rm m}}{4} - (\omega_2+\Delta_1)(\omega_2+\Delta_2) - i\left[ \frac{\gamma_{\rm o}}{2}(\omega_2+\Delta_2) + \frac{\Gamma_{\rm m}}{2}(\omega_2+\Delta_1) \right] } d\omega_2 \nonumber\\
&& = 0,
\end{eqnarray}
\begin{eqnarray}
&&\langle \xi_{ac}^{\dagger}(t) b_{ac}(t) \rangle \nonumber\\
&& =\frac{1}{4\pi^2} \int_{-\infty}^{\infty} e^{i\omega_1 t} d\omega_1 
\int_{-\infty}^{\infty} e^{-i\omega_2 t} \langle \bar{\xi}_{ac}^{\dagger}(\omega_1)\bar{b}_{ac}(\omega_2) \rangle d\omega_2, \nonumber\\
&& = -\frac{1}{4\pi^2} \int_{-\infty}^{\infty} e^{i\omega_1 t} d\omega_1 
\int_{-\infty}^{\infty} e^{-i\omega_2 t}
\frac{ i g_{\rm om}\sqrt{\gamma_{\rm o}}\langle \bar{\xi}_{ac}^{\dagger}(\omega_1)\bar{\xi}_{as}(\omega_2) \rangle   + \left[ i(\omega_2+\Delta_1)-\frac{\gamma_{\rm o}}{2} \right]\sqrt{\Gamma_{\rm m}}\langle \bar{\xi}_{ac}^{\dagger}(\omega_1)\bar{\xi}_{ac}(\omega_2) \rangle  }
{ g_{\rm om}^2 + \frac{\gamma_{\rm o}\Gamma_{\rm m}}{4} - (\omega_2+\Delta_1)(\omega_2+\Delta_2) - i\left[ \frac{\gamma_{\rm o}}{2}(\omega_2+\Delta_2) + \frac{\Gamma_{\rm m}}{2}(\omega_2+\Delta_1) \right] } d\omega_2 \nonumber\\
&& = -\frac{\sqrt{\Gamma_{\rm m}}n_{th}}{2\pi}
\int_{-\infty}^{\infty} \frac{ i(\omega+\Delta_1) - \frac{\gamma_{\rm o}}{2} }
{ g_{\rm om}^2 + \frac{\gamma_{\rm o}\Gamma_{\rm m}}{4} - (\omega+\Delta_1)(\omega+\Delta_2) - i\left[ \frac{\gamma_{\rm o}}{2}(\omega+\Delta_2) + \frac{\Gamma_{\rm m}}{2}(\omega+\Delta_1) \right] } d\omega,
\end{eqnarray}
\begin{eqnarray}
&&\langle \xi_{as}^{\dagger}(t)b_{ac}(t) \rangle \nonumber\\
&& = -\frac{1}{4\pi^2} \int_{-\infty}^{\infty} e^{i\omega_1 t} d\omega_1 
\int_{-\infty}^{\infty} e^{-i\omega_2 t}
\frac{ ig_{\rm om}\sqrt{\gamma_{\rm o}}\langle \bar{\xi}_{as}^{\dagger}(\omega_1) \bar{\xi}_{as}(\omega_2) \rangle  + \left[ i(\omega_2+\Delta_1)-\frac{\gamma_{\rm o}}{2} \right]\sqrt{\Gamma_{\rm m}}\langle \bar{\xi}_{as}^{\dagger}(\omega_1)\bar{\xi}_{ac}(\omega_2) \rangle     }
{ g_{\rm om}^2 + \frac{\gamma_{\rm o}\Gamma_{\rm m}}{4} - (\omega_2+\Delta_1)(\omega_2+\Delta_2) - i\left[ \frac{\gamma_{\rm o}}{2}(\omega_2+\Delta_2) + \frac{\Gamma_{\rm m}}{2}(\omega_2+\Delta_1) \right] } d\omega_2 \nonumber\\
&& = 0,
\end{eqnarray}
and
\begin{eqnarray}
&&\langle a_{as}^{\dagger}(t)\xi_{ac}(t) \rangle \nonumber\\
&& = -\frac{1}{4\pi^2} \int_{-\infty}^{\infty} e^{i\omega_1 t} d\omega_1 
\int_{-\infty}^{\infty} e^{-i\omega_2 t}
\frac{ \left[ -i(\omega_1+\Delta_2)-\frac{\Gamma_{\rm m}}{2} \right]\sqrt{\gamma_{\rm o}} \langle \bar{\xi}_{as}^{\dagger}(\omega_1)\bar{\xi}_{ac}(\omega_2) \rangle   - i g_{\rm om}\sqrt{\Gamma_{\rm m}} \langle \bar{\xi}_{ac}^{\dagger}(\omega_1)\bar{\xi}_{ac}(\omega_2) \rangle     }
{ g_{\rm om}^2 + \frac{\gamma_{\rm o}\Gamma_{\rm m}}{4} - (\omega_1+\Delta_1)(\omega_1+\Delta_2) + i\left[ \frac{\gamma_{\rm o}}{2}(\omega_1+\Delta_2) + \frac{\Gamma_{\rm m}}{2}(\omega_1+\Delta_1) \right] } d\omega_2 \nonumber\\
&& = \frac{ig_{\rm om}n_{th}\sqrt{\Gamma_{\rm m}}}{2\pi}
\int_{-\infty}^{\infty} \frac{1}{ g_{\rm om}^2 + \frac{\gamma_{\rm o}\Gamma_{\rm m}}{4} - (\omega+\Delta_1)(\omega+\Delta_2) + i\left[ \frac{\gamma_{\rm o}}{2}(\omega+\Delta_2) + \frac{\Gamma_{\rm m}}{2}(\omega+\Delta_1) \right] }d\omega.
\end{eqnarray}
Thus we have
\begin{eqnarray}
&&\sqrt{\gamma_{\rm o}}\left( \langle \xi_{as}^{\dagger}(t)a_{as}(t) \rangle 
+ \langle \xi_{as}^{\dagger}(t)a_{as}(t) \rangle^{*}  \right) = 0, \nonumber\\
&&\sqrt{\Gamma_{\rm m}} \left( \langle \xi_{ac}^{\dagger}(t) b_{ac}(t) \rangle 
+ \langle \xi_{ac}^{\dagger}(t) b_{ac}(t)\rangle^{*} \right) \nonumber\\
&&=\frac{\Gamma_{\rm m}n_{th}}{2\pi}
\int_{-\infty}^{\infty} \frac{ \gamma_{\rm o}(g_{\rm om}^2+\frac{\gamma_{\rm o}\Gamma_{\rm m}}{4}) + \Gamma_{\rm m}(\omega+\Delta_1)^2 }
{ \left[ g_{\rm om}^2+\frac{\gamma_{\rm o}\Gamma_{\rm m}}{4} - (\omega^2 + (\Delta_1+\Delta_2)\omega+\Delta_1\Delta_2) \right]^2 + \left[ \frac{\gamma_{\rm o}}{2}(\omega+\Delta_2) + \frac{\Gamma_{\rm m}}{2}(\omega+\Delta_1) \right]^2 } d\omega, \nonumber\\
&&\sqrt{\gamma_{\rm o}}\langle \xi_{as}^{\dagger}(t)b_{ac}(t) \rangle
+ \sqrt{\Gamma_{\rm m}} \langle a_{as}^{\dagger}(t)\xi_{ac}(t) \rangle \nonumber\\
&&=\frac{ig_{\rm om}n_{th}\Gamma_{\rm m}}{2\pi}
\int_{-\infty}^{\infty}
\frac{ g_{\rm om}^2+\frac{\gamma_{\rm o}\Gamma_{\rm m}}{4} - (\omega^2 + (\Delta_1+\Delta_2)\omega+\Delta_1\Delta_2)   - i\left[ \frac{\gamma_{\rm o}}{2}(\omega+\Delta_2) + \frac{\Gamma_{\rm m}}{2}(\omega+\Delta_1) \right]    }
{ \left[ g_{\rm om}^2+\frac{\gamma_{\rm o}\Gamma_{\rm m}}{4} - (\omega^2 + (\Delta_1+\Delta_2)\omega+\Delta_1\Delta_2) \right]^2 + \left[ \frac{\gamma_{\rm o}}{2}(\omega+\Delta_2) + \frac{\Gamma_{\rm m}}{2}(\omega+\Delta_1) \right]^2 } d\omega
\end{eqnarray}
By giving values of system parameters in our experiment, the above two integrals can be calculated as follows
\begin{eqnarray}
\int_{-\infty}^{\infty} \frac{ \gamma_{\rm o}(g_{\rm om}^2+\frac{\gamma_{\rm o}\Gamma_{\rm m}}{4}) + \Gamma_{\rm m}(\omega+\Delta_1)^2 }
{ \left[ g_{\rm om}^2+\frac{\gamma_{\rm o}\Gamma_{\rm m}}{4} - (\omega^2 + (\Delta_1+\Delta_2)\omega+\Delta_1\Delta_2) \right]^2 + \left[ \frac{\gamma_{\rm o}}{2}(\omega+\Delta_2) + \frac{\Gamma_{\rm m}}{2}(\omega+\Delta_1) \right]^2 } d\omega
&=&2\pi, \nonumber\\
\int_{-\infty}^{\infty}
\frac{ g_{\rm om}^2+\frac{\gamma_{\rm o}\Gamma_{\rm m}}{4} - (\omega^2 + (\Delta_1+\Delta_2)\omega+\Delta_1\Delta_2)   - i\left[ \frac{\gamma_{\rm o}}{2}(\omega+\Delta_2) + \frac{\Gamma_{\rm m}}{2}(\omega+\Delta_1) \right]    }
{ \left[ g_{\rm om}^2+\frac{\gamma_{\rm o}\Gamma_{\rm m}}{4} - (\omega^2 + (\Delta_1+\Delta_2)\omega+\Delta_1\Delta_2) \right]^2 + \left[ \frac{\gamma_{\rm o}}{2}(\omega+\Delta_2) + \frac{\Gamma_{\rm m}}{2}(\omega+\Delta_1) \right]^2 } d\omega
&=&0.
\end{eqnarray}
Therefore, the dynamics of the mean photon and phonon numbers can be given by
\begin{eqnarray}\label{Motion equation of second order moments without noises}
\frac{d N_a}{d t} &=& -\gamma N_a - i g_{om} ( \langle a_{as}^{\dagger} b_{ac}\rangle - \langle a_{as}^{\dagger} b_{ac} \rangle^* ), \nonumber\\
\frac{d N_b}{d t} &=& -\Gamma N_b + i g_{om} ( \langle a_{as}^{\dagger} b_{ac}\rangle - \langle a_{as}^{\dagger} b_{ac} \rangle^* )
+ \Gamma_{\rm m}n_{th}, \nonumber\\
\frac{d \langle a_{as}^{\dagger}b_{ac}\rangle}{d t} &=& - \left (i(\Delta_1-\Delta_2)+\frac{\gamma+\Gamma}{2}\right )
\langle a_{as}^{\dagger}b_{ac}\rangle - i g_{om} N_a + i g_{om} N_b.
\end{eqnarray}
Here we consider the Brillouin cooling at steady-state in our experiment. Thus the phonon occupation in the stable regime can be given by
\begin{eqnarray}
N_b^{\rm ss} = \frac{ 4g_{\rm om}^2(\Gamma_{\rm m}+\gamma_{\rm o}) + \gamma_{\rm o}(\Gamma_{\rm m}+\gamma_{\rm o})^2 + 4\gamma_{\rm o}(\Delta_1-\Delta_2)^2 }
{ 4g_{\rm om}^2(\Gamma_{\rm m}+\gamma_{\rm o}) + \gamma_{\rm o}\Gamma_{\rm m}(\Gamma_{\rm m}+\gamma_{\rm o}) + 4\gamma_{\rm o}\Gamma_{\rm m}(\Delta_1-\Delta_2)^2/(\Gamma_{\rm m}+\gamma_{\rm o}) } \cdot \frac{\Gamma_{\rm m}}{\Gamma_{\rm m}+\gamma_{\rm o}} n_{th}.
\end{eqnarray}

\begin{figure}[h]
    \centering
    \includegraphics[width=10 cm, clip]{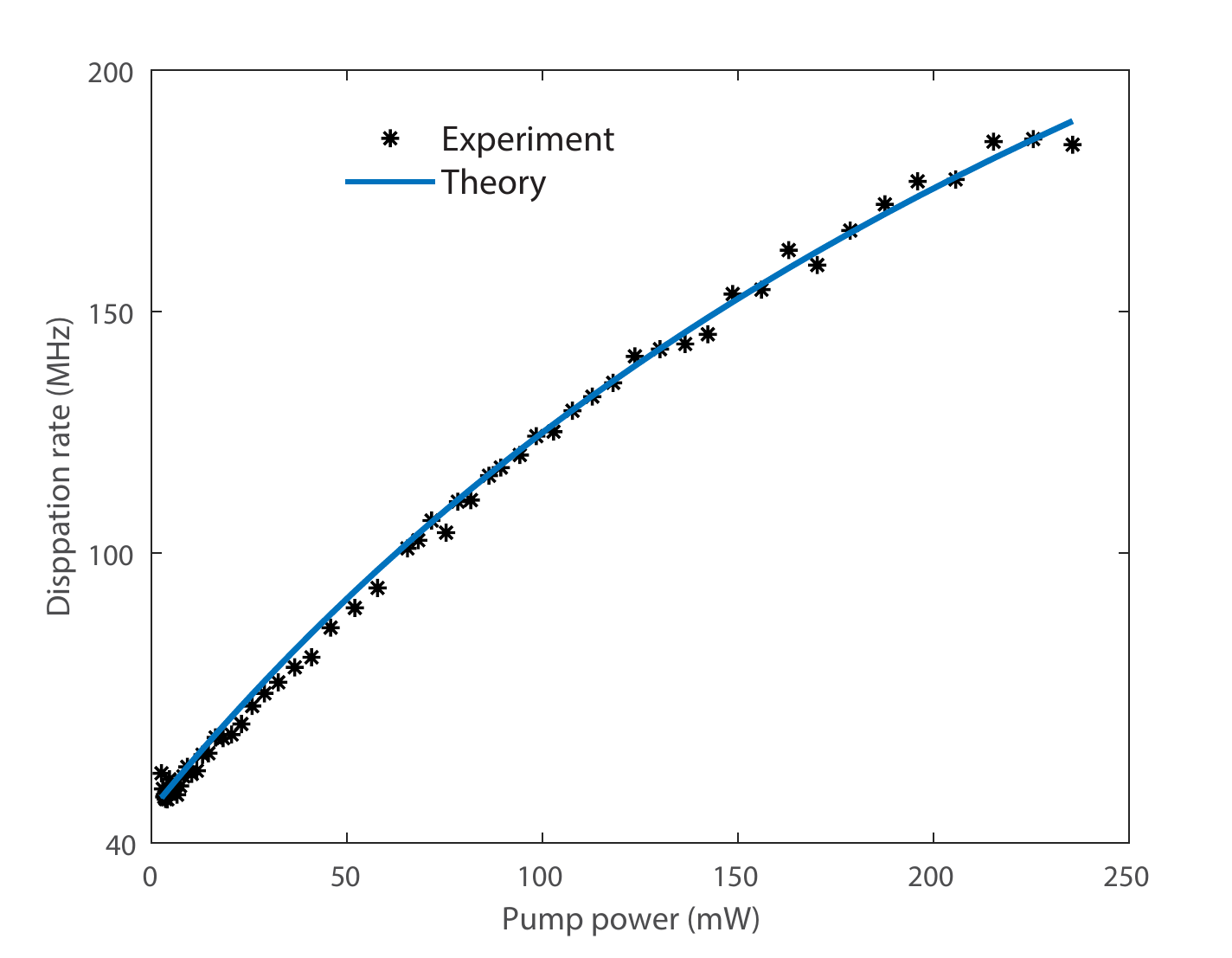}
    \caption{(Color online). Linewidth of the anti-Stokes spectrum versus pump power.}
    \label{figA1}
\end{figure}

We are interested in the case where the anti-Stokes mode and the acoustic mode are phase-matched with the pump wave, i.e., $\Delta_1=\Delta_2=0$. In this case, the phonon occupation at the steady state can be expressed as
\begin{eqnarray}
N_b^{\rm ss} = \frac{ 4g_{\rm om}^2 + \gamma_{\rm o}(\Gamma_{\rm m}+\gamma_{\rm o})  }
{ 4g_{\rm om}^2 + \gamma_{\rm o}\Gamma_{\rm m}  } \cdot \frac{\Gamma_{\rm m}}{\Gamma_{\rm m}+\gamma_{\rm o}} n_{th}.
\end{eqnarray}

Actually, for the anti-Stokes process, the anti-Stokes wave can introduce extra damping for the acoustic wave by absorbing acoustic phonons, which causes the broadening of the linewidth of the acoustic mode. The effective damping rate of the cooled acoustic mode is associated with the cooling rate as follows
\begin{eqnarray}
\frac{\Gamma_{\rm m}}{\Gamma_{\rm eff}} = \frac{N_{b}^{ss}}{n_{th}},
\end{eqnarray}
where we assume that the initial state of the acoustic mode is the thermal state. Thus the effective acoustic damping rate can be given by
\begin{eqnarray}\label{gamma effective}
\Gamma_{\rm eff} = \Gamma_{\rm m} + \frac{4g_{\rm om}^2\gamma_{\rm o}}{ 4g_{\rm om}^2 + \gamma_{\rm o}(\Gamma_{\rm m}+\gamma_{\rm o}) }.
\end{eqnarray}
Given the system parameters measured in our experiment: $\Gamma_{\rm m}=46.8~$MHz, $\gamma_{\rm o}=364~$MHz, $c=3\times10^8~$m/s, $n=2.5$, $L=0.5~$m, and $G_{\rm B}=164~\rm{m^{-1}W^{-1}}$, we show the linewidth of the acoustic mode with respect to pump power in Fig.~\ref{figA1}, where the blue-solid curve denotes the effective acoustic damping rate calculated in Eq.~(\ref{gamma effective}).
Here the optomechanical coupling strength can be evaluated as $g_{\rm om} = \sqrt{G_{\rm B}\Gamma_{\rm m} P L c/(4n)}$, where $c$ is the speed of light and $n$ is the refractive index.

\begin{figure}
\centering
\includegraphics{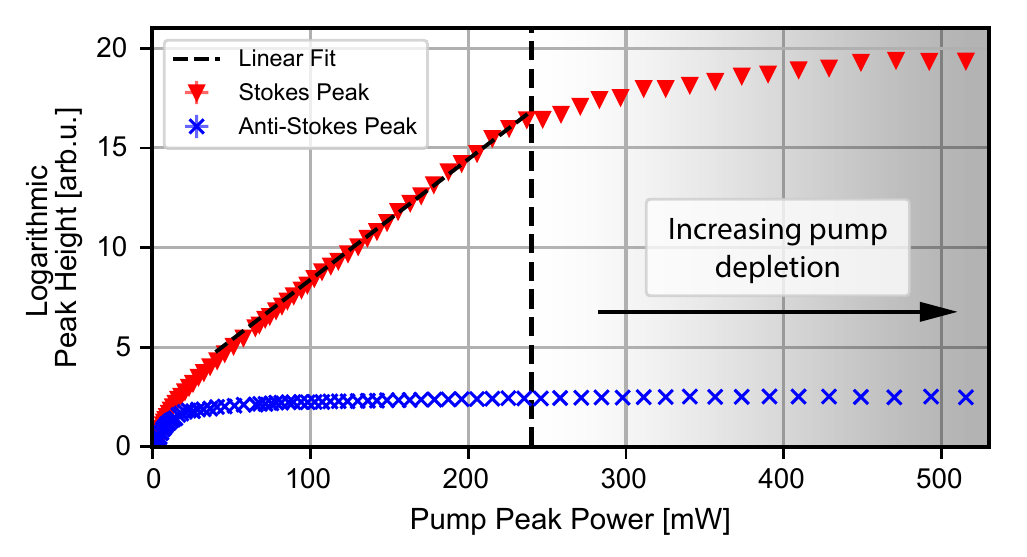}
\caption{(Color online). Peak height of the Stokes (red triangles) and anti-Stokes (blue
crosses) resonances in logarithmic scale as a function of pump power. The vertical dashed line indicates the threshold for pump depletion, in which the Stokes peak no longer increases logarithmically.}
\label{figA2}
\end{figure}  

\begin{figure}
\centering
\includegraphics{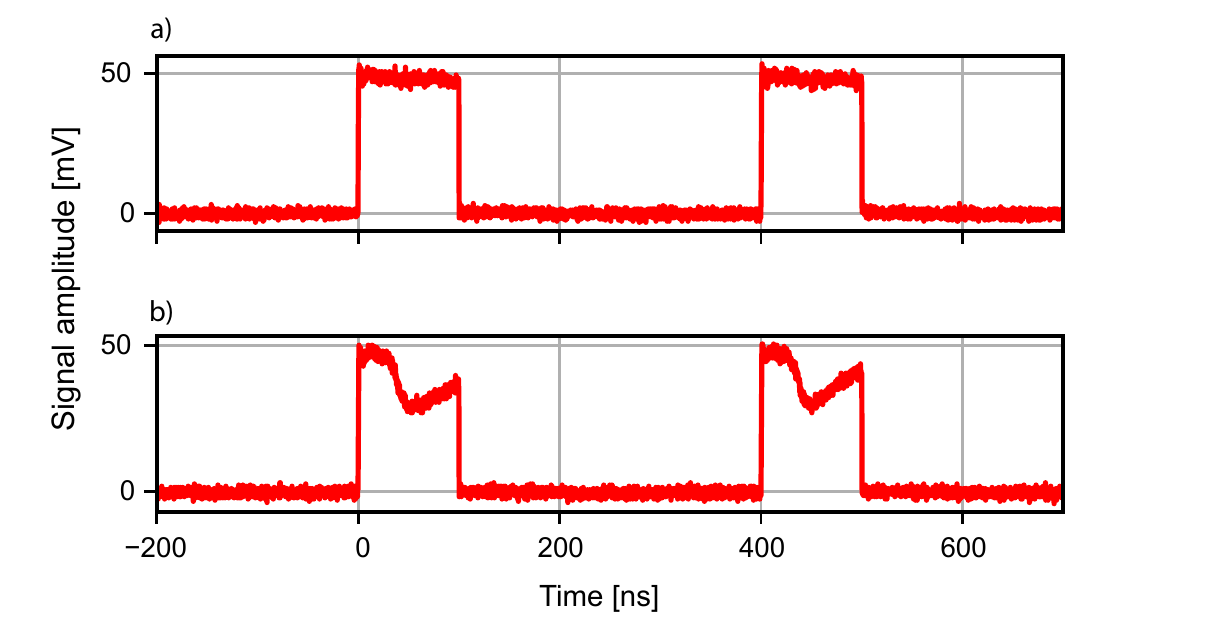}
\caption{ Shape of transmitted optical pump pulses after the tapered chalcogenide PCF. Square pulse with a length of 100 ns, 25\% duty cycle. a) 100 mW peak pump power, below pump depletion threshold. b) 500 mW peak pump power, above pump depletion threshold.}
\label{figA3}
\end{figure}  

\section{Pump depletion limiting cooling}\label{S3}

Experimentally, it is observed that the lowest phonon mode effective temperature achieved is 74 K, higher than the minimum temperature expected from the theory (around 36 K). This is caused by a depletion of the pump via the strong Stokes wave. All the theoretical derivations in this paper are made under the assumption of undepleted pump, i.e. the amount of energy scattered in the Brillouin-Mandelstam process being negligible compared to the incident pump energy. Given the extremely high gain of the sample, this assumption was observed to not apply to the whole measurement range.

As shown in Eq. (\ref{gamma effective}), higher pump power results in a higher degree of cooling. On the other hand, higher pump power will also result in a stronger Stokes response from the system \cite{boydNonlinearOptics2020_supp}. As the Stokes response increases exponentially with input power, after a certain threshold, the amount of Stokes light will start to deplete the pump wave \cite{ippenStimulatedBrillouinScattering2003_supp}. This is shown in Fig. \ref{figA2}. For pump powers higher than 240 mW, the height of the Stokes resonance stops increasing exponentially, indicating the start of the pump depletion regime. This depletion of the pump can also be observed in the pump signal transmitted through the sample (Fig. \ref{figA3}). Below the depletion threshold, the pump pulses maintain their initial square shape. As the power is increased, a dip in the amplitude can be observed after the sample.
This depletion of the pump via the Stokes process limits the maximum cooling power achievable from the system. As more pump power is scattered into the Stokes wave, less energy is available to further cool the anti-Stokes resonant phonons.


%

\end{document}